\documentclass[usenatbib,usegraphicx]{mn2e}

\usepackage{graphicx}
\usepackage{url}
\usepackage{textcomp} 

\font \bolditalics = cmmib10
\def\bx#1{\leavevmode\thinspace\hbox{\vrule\vtop{\vbox{\hrule\kern1pt
        \hbox{\vphantom{\tt/}\thinspace{\bf#1}\thinspace}}
      \kern1pt\hrule}\vrule}\thinspace}
\def \vc #1{{\textfont1=\bolditalics \hbox{$\bf#1$}}}

\def\eg{{\bf e}}
\def\kg{{\bf k}}

\def\ggr{{\bf g}}
\def\eg{{\bf e}}
\def\sg{{\bf s}}

\def\thetavc{{\vc \theta}}

\def\gammag{{\vc \gamma}}

\title[CFHTLenS: Mapping the Large Scale Structure]{CFHTLenS: Mapping the Large Scale Structure with Gravitational Lensing}

\author[L.~Van Waerbeke et al.]{L.~Van Waerbeke$^{1}$\thanks{Email: waerbeke@phas.ubc.ca}, J. Benjamin$^{1}$, T. Erben$^2$, C. Heymans$^3$, H. Hildebrandt$^{2,1}$,
\newauthor H. Hoekstra$^{5,6}$, T.D. Kitching$^{4,3}$, Y. Mellier$^{7,8}$, L. Miller$^9$, J. Coupon$^{10}$,
\newauthor J. Harnois-D{\'e}raps$^{1,11,12}$, L. Fu$^{13}$, M. Hudson$^{14,15}$, M. Kilbinger$^{16,7}$,
\newauthor K. Kuijken$^5$, B. Rowe$^{17,18}$, T. Schrabback$^{2,5,19}$, E. Semboloni$^5$, S. Vafaei$^1$,
\newauthor E. van Uitert$^{5,2}$, M. Velander$^{9,5}$
\\
\\
$^1$University of British Columbia, Department of Physics and Astronomy, 6224 Agricultural Road, Vancouver, BC V6T 1Z1, Canada\\
$^2$Argelander-Institut f\"ur Astronomie, Auf dem H\"ugel 71, 53121 Bonn, Germany\\
$^3$Scottish Universities Physics Alliance, Institute for Astronomy, University of Edinburgh, Royal Observatory, Blackford Hill,\\ Edinburgh EH9 3HJ, UK\\
$^4$Mullard Space Science Laboratory, University College London, Holmbury St Mary, Dorking, Surrey RH5 6NT, UK\\
$^5$Leiden Observatory, Leiden University, Niels Bohrweg 2, 2333 CA Leiden, The Netherlands\\
$^6$Department of Physics and Astronomy, University of Victoria, Victoria, BC V8P 5C2, Canada\\
$^7$Institut d'Astrophysique de Paris, Universit{\'e} Pierre et Marie Curie - Paris 6, 98 bis Boulevard Arago, F-65014 Paris, France\\
$^7$Institut d'Astrophysique de Paris, CNRS, UMR 7095, 98 bis Boulevard Arago, F-75014 Paris, France\\
$^9$Department of Physics, Oxford University, Keble Road, Oxford OX1 3RH, UK\\
$^{10}$Institute of Astronomy and Astrophysics, Academia Sinica, P.O. Box 23-141, Taipei 10617, Taiwan\\
$^{11}$Canadian Institute for Theoretical Astrophysics, University of Toronto, M5S 3H8, Ontario, Canada\\
$^{12}$Department of Physics, University of Toronto, M5S 1A7, Ontario, Canada\\
$^{13}$Key Lab for Astrophysics, Shanghai Normal University, 100 Guilin Road, 200234, Shanghai, PR China\\
$^{14}$Department of Physics and Astronomy, University of Waterloo, Waterloo, ON N2L 3G1, Canada\\
$^{15}$Perimeter Institute for Theoretical Physics, 31 Caroline St N, Waterloo, Ontario, N2L 2Y5, Canada\\
$^{16}$CEA Saclay, Service d'Astrophysique (SAp), Orme des Merisiers, B\^at 709, F-91191 Gif-sur-Yvette, France\\
$^{17}$Department of Physics and Astronomy, University College London, Gower Street, London WC1E 6BT, UK\\
$^{18}$California Institute of Technology, 1200 E California Boulevard, Pasadena CA 91125, USA\\
$^{19}$Kavli Institute for Particle Astrophysics and Cosmology, Stanford University, 382 Via Pueblo Mall, Stanford, CA 94305-4060, USA\\
}

\date{Released 2012 Xxxxx XX}

\pagerange{\pageref{firstpage}--\pageref{lastpage}} \pubyear{2012}

\include{journal_defs}

\begin{document}
\label{firstpage}

\maketitle

\newpage
\begin{abstract}
We present a quantitative analysis of the largest contiguous maps of projected mass density obtained from gravitational lensing shear. We use data from the $154$ deg$^2$ covered by the Canada-France-Hawaii Telescope Lensing Survey (CFHTLenS). Our study is the first attempt to quantitatively characterize the scientific value of lensing maps, which could serve in the future as a complementary approach to the study of the dark universe with gravitational lensing. We show that mass maps contain unique cosmological information beyond that of traditional two-points statistical analysis techniques.

Using a series of numerical simulations, we first show how, reproducing the CFHTLenS observing conditions, gravitational lensing inversion provides a reliable estimate of the projected matter distribution of large scale structure. We validate our analysis by quantifying the robustness of the maps with various statistical estimators. We then apply the same process to the CFHTLenS data. We find that the 2-points correlation function of the projected mass is consistent with the cosmological analysis performed on the shear correlation function discussed in the CFHTLenS companion papers. The maps also lead to a significant measurement of the third order moment of the projected mass, which is in agreement with analytic predictions, and to a marginal detection of the fourth order moment. Tests for residual systematics are found to be consistent with zero for the statistical estimators we used.
A new approach for the comparison of the reconstructed mass map to that predicted from the galaxy distribution reveals the existence of giant voids in the dark matter maps as large as $3$ degrees on the sky. Our analysis shows that lensing mass maps are not only consistent with the results obtained by the traditional shear approach, but they also appear promising for new techniques such as peak statistics and the morphological analysis of the projected dark matter distribution.
\end{abstract}
\begin{keywords}
dark matter - large-scale structure of the Universe - gravitational lensing
\end{keywords}

\section{Introduction}
\label{sec:intro}

Gravitational lensing is a powerful tool for the study of the dark matter distribution in the Universe. The statistical analysis of the distortion and magnification of lensed galaxies provides unique information on structure formation processes. For example, two point statistics of the shear or convergence can be used to constrain the dark matter power spectrum, the growth of structure, and cosmological parameters \citep[see][for recent reviews]{Munshi08,2008ARNPS..58...99H}. Over the past ten years, multiple groups have reported improved constraints on the mass density parameter and the power spectrum normalisation using the shear correlation function. However, it is well known that gravitational lensing contains a lot more information than the amplitude and shape of the mass power spectrum. The distortion (shear), in particular, can be used to reconstruct the projected mass density, or mass maps, thus making the information on the distribution of dark matter available in a different form. The use of mass maps as a cosmological tool has received little attention so far.

The purpose of this paper is to explore the extent to which mass maps can access the cosmological information not captured by the two points statistics. Several groups have explored different theoretical routes beyond two points statistics. For instance, higher order shear measurements are a sensitive measure of the gravitational collapse process through mode coupling in the non-linear regime \citep{1997A&A...322....1B,2001MNRAS.322..918V,2003MNRAS.344..857T,2005A&A...442...69K}, and can also be used as an indicator of non-Gaussian non-linearity in the primordial dark matter distribution \citep{2004MNRAS.348..897T,2005MNRAS.356..386V}. More exotic statistical estimators involve global statistical tools, such as the Minkowski functional, as opposed to {\it local} measurements based on the shear correlation functions. Morphology of large scale structures \citep{1994A&A...288..697M} is also a probe of the non-linear processes in action during structure formation \citep{2001ApJ...551L...5S}.

The peak statistics is also an estimator that can be used to probe the number of dark matter haloes as a function of redshift and mass \citep{2000MNRAS.313..524V,2000ApJ...530L...1J,2010PhRvD..81d3519K,2010A&A...519A..23M}, and it can potentially constrain the halo mass function \citep{2011PhRvD..84d3529Y}. The effect of the large scale dark matter environment on the galaxy-galaxy lensing signal could be investigated; for instance mass maps could be used to quantify to what extent specific galaxy properties (e.g. star formation rate, dust content, stellar population) depend on the global dark matter environment such as a supercluster or a giant void.

Most of these global statistics, like peaks and structure morphology, are non-local, and therefore cannot be expressed as a combination of a subset of moments of the shear, which are, by definition, {\it local} measurements. The study of global features of the mass distribution require the mapping of the full 2-dimensional mass. Dark matter maps can also be used to understand global features in the baryon/dark matter relation via cross-correlation with maps from different wavelengths surveys, whether Sunyaev-Zel{\textquotesingle}dovich, X-ray, or atomic hydrogen. The use of mass maps is therefore complementary to the analysis of the shear and magnification two-point correlation functions. This motivates the need to have reliable mass maps from gravitational lensing data.

Mass reconstruction is a common tool for the study of galaxy clusters \citep{2000ApJ...532...88H,2006ApJ...648L.109C,2006A&A...451..395C,2008MNRAS.385.1431H}; it is mostly used to evaluate qualitatively the matching between the baryonic and dark matter distributions, and to explore the possible existence of dark clumps. A few rare cases suggest their possible existence \citep{2000A&A...355...23E,2007ApJ...668..806M,2012ApJ...747...96J}, although the significance of these detections is currently unclear. Mass reconstruction in cluster environments concern regions of relatively high lensing signal and small angular scale.
Mass reconstruction of weaker lensing signal, beyond galaxy cluster scales, was performed in \cite{2007Natur.445..286M} with the $1.64$ deg$^2$ area of the COSMOS survey. However, our work shows that sampling variance is still important at even larger scales, implying that a relatively {\it small} survey like COSMOS does not provide a fair sample of the dark matter distribution.

In this paper we use the $154$ deg$^2$ CFHTLenS lensing data \citep{2012arXiv1210.8156E,2012MNRAS.427..146H,2013MNRAS.429.2858M,2012MNRAS.421.2355H}
to perform lensing mass reconstruction over angular scales of several degrees. We do not attempt to perform a 3-dimensional mass reconstruction
 \citep{2007Natur.445..286M,2012MNRAS.419..998S,2011ApJ...727..118V,2012A&A...539A..85L} because such reconstruction using ground-based data would only yield meaningful results for galaxy clusters relatively close in redshift, $z<0.3$, or for very massive and rare clusters at higher redshift, $z<0.6$ \citep{2009MNRAS.399...48S}. We explore the measurement of convergence statistics from gravitational lensing maps and the connection between dark and baryonic matter up to a few degrees across. In particular, we focus on the importance of the effect of noise in the reconstruction, and how to account for it in the interpretation of the measurements. To this end, we use ray-tracing simulations to test our reconstruction and analysis procedure.

This paper is organized as follows. Notation and definitions are described in Section 2. The analysis of simulated reconstructed mass maps is discussed in Section 3. In Section 4, we present the results from the CFHTLenS data and our new approach for the comparison of dark matter and baryonic matter. We present our conclutions in Section 6.

\section{Map making and cosmology}

\subsection{Mass reconstruction}

Mass maps are proportional to the projected mass convolved with the lensing kernel. They can be constructed from the shear measurement. A galaxy at position $\thetavc=(\theta_1,\theta_2)$ on the sky is characterised by its redshift $z$ and its shear components $\gamma_i(\thetavc)$, where $i=1,2$. In this paper, we use the flat sky tangent plane approximation. The relation between $\gammag$ and the convergence $\kappa$ involves the gravitational deflection potential $\Psi(\thetavc)$, defined as:

\begin{equation}
\gamma_i(\thetavc)={\rm G}_i \Psi(\thetavc),
\label{gammadef}
\end{equation}
where ${\rm G}_1=\partial_{11}-\partial_{22}$ and ${\rm G}_2=2\partial_{12}$. The function $\Psi$ is given by a line-of-sight integral
of the 3-dimensional matter gravitational potential $\Phi$ via:

\begin{equation}
\Psi(\thetavc)=\int_0^{w_s} {\rm d} w' {f_K(w-w')\over f_K(w') f_K(w)} \Phi(f_K(w')\thetavc,w').
\label{potential}
\end{equation}
The comoving radial distance at redshift $z$ is given by $w(z)$, and $f_K(w)$ is the corresponding angular diameter distance. The comoving distance at the source redshift is $w_s$. The 3-dimensional gravitational potential depends on the mass density contrast $\delta$ via the Poisson equation:

\begin{equation}
\Delta_{3D}\Phi={3 H_0^2\Omega_0\over 2 a}\delta,
\label{phidef}
\end{equation}
where $H_0$ is the Hubble constant, $\Omega_0$ is the matter density, and $a$ is the scale factor. The convergence $\kappa(\thetavc)$ is defined as the line-of-sight projection of the 2-dimensional Laplacian of $\psi$:

\begin{equation}
\kappa(\thetavc)=\Delta\Psi(\thetavc),
\label{kappadef}
\end{equation}
where $\Delta=\partial_{11}+\partial_{22}$, and the derivative $\partial_{33}$ of the gravitational potential along the line of sight vanishes on average due
to the Limber approximation \citep{1998ApJ...498...26K}. The observable quantity is the reduced shear $\ggr$:

\begin{equation}
\ggr={\gammag\over 1-\kappa}.
\label{gdef}
\end{equation}
Due to the intrinsic ellipticity of galaxies, galaxy shapes provide a noisy estimate of $\gammag$. This estimator, $\eg^{\rm obs}$, is given by:

\begin{equation}
\eg^{\rm obs}={\ggr+\eg^{\rm int}\over 1-\ggr\eg^{\rm int}}.
\label{gdef2}
\end{equation}
The intrinsic ellipticity $\eg^{\rm int}$ is a generic term that contains a contribution from the intrinsic galaxy shape and from the measurement noise. The ensemble average $\langle \eg^{\rm obs}\rangle$ is an unbiased estimator of the reduced shear $\ggr$ \citep{1988A&A...191...39K}. In the case of CFHTLenS, a complete description of the measurement noise is given in \cite{2012MNRAS.427..146H} and \cite{2013MNRAS.429.2858M}.

Weak lensing studies assume that both the shear and convergence amplitudes are much smaller than unity, ($|\gammag|, \kappa \ll 1$) so that one can use the weak lensing approximation ($\ggr\simeq\gammag$) for mass reconstruction. In that case, Eq.(\ref{gdef2}) reduces to $\eg^{\rm obs}\simeq \gammag+\eg^{\rm int}$, and it follows that by comparing Eq.(\ref{gammadef}) and Eq.(\ref{kappadef}), one can reconstruct the convergence from the observed shear in Fourier space \citep{1993ApJ...404..441K}.
For the weak lensing approximation, we first write the relation between the shear and convergence Fourier components
$\hat\kappa(\sg)$ and $\hat\gammag(\sg)$, where we define
the wavevector $\sg=(s_1,s_2)=2\pi/\thetavc$ as the 2-dimensional analogue of $\thetavc$ in Fourier space:

\begin{equation}
\hat\kappa(\sg)={1\over 2}\left({k_1^2-k_2^2\over k_1^2+k_2^2}\right)\hat\gamma_1(\sg)+{k_1k_2\over k_1^2+k_2^2}\hat\gamma_2(\sg).
\label{ks93def}
\end{equation}
Following \cite{1993ApJ...404..441K}, the shear data are first regularised with a smoothing window and then Eq.(\ref{ks93def}) evaluated from the smoothed Fourier components. Since real data also contain masked regions ( e.g. bright stars and sometimes area missing from the detector), the smoothing takes into account the number of pixels being masked within each smoothing window, so that the resulting averaged quantity in that window is not biased. The practical procedure for obtaining a mass map from reduced shear is as follows:

\begin{enumerate}
\item The data $\eg^{\rm obs}(\thetavc_{ij})=\eg^{\rm obs}_{ij}$ are first placed on a regular grid $\thetavc_{ij}$ and then smoothed. At pixel location $\thetavc_{ij}$, the
smoothed ellipticity $\overline{\eg}_{ij}$ is given by:

\begin{equation}
\overline{\eg}_{ij}={\sum_{kl} W_{\theta_0}(\theta_{kl}) w(\theta_{i-k;j-l})\eg^{\rm obs}_{i-k;j-l}\over \sum_{kl} w(\theta_{i-k;j-l})}\label{gsmooth},
\end{equation}
where $W_{\theta_0}(\thetavc)$ is a normalised Gaussian smoothing window:

\begin{equation}
W_{\theta_0}(\thetavc)={1\over \pi \theta_0^2} \exp\left(-{|\thetavc|^2\over 2\theta_0^2}\right),
\label{Wgauss}
\end{equation}
and $w(\thetavc)$ is the weight associated with the measurement noise of $\eg(\thetavc)$. Details on the weight specific to CFHTLenS data are given in Section 4.2.
\item The reconstructed mass is obtained from Eq.(\ref{ks93def}). An absence of galaxies should result in a pixel value of zero. However, note that since the Gaussian filter has infinite spatial extension, there are no pixels in the final grid with a value of zero. There are, however, pixels with higher noise (fewer objects) than others, especially when a mask overlaps with the central region of the filter. The noisiest pixels in the reconstructed map are removed if the effective {\it filling factor} within the Gaussian window is below $50$ percent. This point is discussed later in Section 3.2, where the noise in the reconstructed maps plays an important role in the cosmic statistics measurement from the convergence.
\end{enumerate}

The alternative to the \cite{1993ApJ...404..441K} mass reconstruction approach is the full non-linear mass reconstruction as described in \citet{1996ApJ...464L.115B}. A detailed comparison between the two methods is left for a future study. For the purposes of this paper, the KS93 reconstruction is sufficient because we are not trying to recover the lensing signal in the non-linear regime at sub-arcminute scales. We note, however, that the ability of the non-linear mass reconstruction to recover large scale structure statistics was demonstrated in \citet{1999A&A...342...15V}. Therefore, moving to a full non-linear reconstruction does not represent a conceptual challenge.

As shown in \cite{Crittenden02}, the shear can be split into $E$ and $B$ modes, which correspond to the curl-free and curl shear components respectively. In the absence of residual systematics, the scalar nature of the gravitational potential leads to a vanishing $B$-mode. As shown in \cite{Schneider98}, the split between $E$ and $B$ modes is performed by applying the transformation $(\gamma_1,\gamma_2)$ to $(-\gamma_2,\gamma_1)$, which is the same as rotating each galaxy by $45$ degrees. Our mass maps are reconstructed using the $E$ and $B$ modes and the corresponding convergence is called $\kappa_E$ and $\kappa_B$. As we will see later on, it will be necessary to distinguish between the reconstructed (noisy) map, and the underlying true convergence value. We therefore introduce three additional convergence terms. We call $\kappa_{\rm obs}$ the reconstructed convergence such that:

\begin{equation}
\kappa_{\rm obs}=\kappa_E+\kappa_{\rm ran},
\label{Kobs}
\end{equation}
where $\kappa_E$ is the true underlying signal and $\kappa_{\rm ran}$ is the reconstruction noise. This equation is the convergence equivalent of Eq.(\ref{gdef2}) in the weak lensing regime. We also introduce the convergence $\kappa_\perp$, reconstructed from the galaxies rotated by $45$ degrees such that:

\begin{equation}
\kappa_\perp=\kappa_B+\kappa_{\rm ran},
\label{Kperp}
\end{equation}
where $\kappa_B$ should be consistent with zero if the residual systematics are negligible. The validity of this statement will be verified statistically.

\subsection{Cosmology}

The noise-free convergence map $\kappa_{\theta_0}(\thetavc)$, smoothed with a Gaussian filter of size $\theta_0$, can be expressed in terms of the 3-dimensional mass density contrast $\delta(f_K(w)(\thetavc),w)$:

\begin{eqnarray}
\kappa_{\theta_0}(\thetavc)&=&{3\over 2}\Omega_0\int{\rm d}\thetavc'\int_0^{w_H} {\rm d} w {g(w)\over a(w)} \times \cr
&&\delta(f_K(w)(\thetavc-\thetavc'),w) W_{\theta_0}(\thetavc'),
\label{kappamapdef}
\end{eqnarray}
where $W_{\theta_0}(\thetavc)$ is the normalised Gaussian window given by Eq.(\ref{Wgauss}) and $w_H$ represents the radial distance at infinite redshift.
The function $g(w)$ accounts for the distribution of source redshifts:

\begin{equation}
g(w)=\int_w^{w_H}{\rm d}w' p_S(w'){f_K(w'-w)\over f_K(w')},
\end{equation}
where $p_S(w(z))$ is the source redshift distribution.
Eq.(\ref{kappamapdef}) represents the convergence map obtained from the reconstruction process after the shear data have been pixelated and smoothed as described in Section 2. Therefore, the positions on the sky $\thetavc$ are pixels on the regular grid where the map has been computed. The remainder of this paper will focus on sub-degree scales where the flat sky approximation applies. However, we will use the exact full sky formalism for all the lensing quantities, which has the advantage of providing a robust numerical integration via a direct, one pass, summation of the $l$ wavevectors. Furthermore, a conservative cut at $l_{\rm max}=100000$ is used for both the second and third order statistics predictions. The second order moment of the convergence $\langle \kappa^2(\theta_0)\rangle$ can be expressed from the convergence power spectrum $C_l^\kappa$:

\begin{equation}
\langle \kappa^2\rangle_{\theta_0}={1\over 4\pi}\sum_l (2l+1)C_l^\kappa W^2_l(\theta_0),
\label{kappa2def}
\end{equation}
where $W_l(\theta_0)$ are the multipole moments of the smoothing window. For the Gaussian window given by Eq.(\ref{Wgauss}), the multipole moments are given by $W_l(\theta_0)=\exp(-l^2\theta_0^2/4)$. The power spectrum $C_l^\kappa$ can be derived from Eq.(\ref{kappamapdef}), assuming that the small angle and Limber approximations apply \citep{1991ApJ...380....1M,Kaiser92}:

\begin{eqnarray}
C^\kappa_l&=&{9\over 4}\Omega_0^2\int_0^{w_s}{\rm d}w{g^2(w)\over a^2(w)}P_{3D}\left({l\over f_K(w)}; w\right) \cr
&\times & {f_K(w_s-w)f_K(w)\over f_K(w_s)}.
\end{eqnarray}
The 3-dimensional mass power spectrum is defined as:

\begin{equation}
\langle \tilde\delta^*(\kg,w)\tilde\delta(\kg',w)\rangle= (2\pi)^3 \delta_D(\kg-\kg') P_{3D}\left(k; w\right).
\end{equation}
The third order moment $\langle \kappa^3\rangle_{\theta_0}$ can also be expressed analytically using the convergence angular bi-spectrum $B_{l_1 l_2 l_3}^\kappa$, the smoothing filter multipoles $W_l(\theta_0)$, and the Wigner 3-j symbols:

\begin{eqnarray}
\langle \kappa^3\rangle_{\theta_0} &=&{1\over 4\pi}\sum_{l_1 l_2 l_3} \sqrt{(2l_1+1)(2l_2+1)(2l_3+1)\over 4\pi}\cr
&\times &\left(\matrix{l_1 & l_2 & l_3 \cr 0 & 0 & 0}\right) \cr
&\times & B_{l_1 l_2 l_3}^\kappa W_{l_1}(\theta_0) W_{l_2}(\theta_0) W_{l_3}(\theta_0),\nonumber
\label{kappa3def}
\end{eqnarray}
with the angular bi-spectrum given by:

\begin{eqnarray}
B_{l_1 l_2 l_3}^\kappa&=&{81\Omega_0^3\over 8\pi}\sqrt{(2l_1+1)(2l_2+1)(2l_3+1)\over 4\pi} \left(\matrix{l_1 & l_2 & l_3 \cr 0 & 0 & 0}\right)\cr
&\times & \int_0^{w_S} {\rm d}w {g^3(w)\over a^3(w) f_K(w)} \cr
&\times & B_{3D}\left({\l_1\over f_K(w)},{\l_2\over f_K(w)},{\l_3\over f_K(w)};w\right),
\end{eqnarray}
where $B_{3D}$ is the three dimensional bi-spectrum. $B_{3D}$ is calculated in the non-linear regime using non-linear extensions of perturbation theory. Expressions for gravitational lensing are derived in \cite{2001MNRAS.322..918V}, based on the work of \cite{2001MNRAS.325.1312S}.

\section{Simulations}

\subsection{The CFHTL\lowercase{en}S data}

The purpose of the following Section is to validate our mass reconstruction approach using mock catalogues that replicate the true CFHTLenS observing conditions.
 CFHTLenS spans a total survey area of $154$ deg$^2$, covered with a mosaic of $171$ individual pointings observed by the one square degree imager at the Canada-France-Hawaii Telescope. The survey consists of four compact regions called W1, W2, W3 and W4, which cover approximately $72$, $36$, $50$ and $25$ deg$^2$ respectively. Details on the data reduction are described in \cite{2012arXiv1210.8156E}. The effective area is reduced to $120$ deg$^2$ by the masking of bright stars, artificial and natural moving objects, and faulty CCD rows. The observations in the five bands $u^{*}g'r'i'z'$ of the survey allow for the precise measurement of photometric redshifts \citep{2012MNRAS.421.2355H}. The shape measurement with {\it lens}fit is described in detail in \cite{2013MNRAS.429.2858M}.

\subsection{Mock catalogues}

The mock catalogues are constructed from a mixture of real data (the noise, galaxy position, and masking structure of CFHTLenS) and simulated data (the shear signal from N-body simulations). The N-body simulations used in this work are described in \cite{2012MNRAS.426.1262H}. The procedure for generating mock catalogues is as follows:

\begin{figure}
 \begin{center}
   \includegraphics[width = 8.5cm]{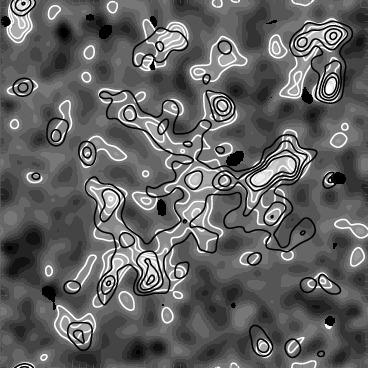}
   \caption{Mass reconstruction for one simulated line of sight covering $12$ square degrees. The continuous background map with white contours represents the reconstructed lensing mass (convergence) with masks shown by the black regions. The white contours show the $1$, $2$, $3$, and $4$ $\sigma$ contours, while the black contours show the $1$, $2$, $3$, and $4$ $\sigma$ levels in the noise-free map. $\sigma$ is the convergence rms measured on the mass reconstruction.}
\label{simul1}
\end{center}
\end{figure}

\begin{enumerate}

\item Projected shear and convergence maps are constructed from the combination of redshift slices for all source redshifts available in the simulation\footnote{The source redshifts are $0.025, 0.075, 0.126, 0.178, 0.232, 0.287, 0.344,$ $0.402, 0.463, 0.526, 0.591, 0.659, 0.73, 0.804, 0.881, 0.961, 1.071, 1.215,$ $ 1.371, 1.542, 1.728, 1.933, 2.159, 2.411, 2.691, 3.004$.}. The maps are sampled on a $1024\times 1024$ grid and cover $12.4$ deg$^2$ each, which corresponds to a native pixel size of $0.21$ arcmin. These maps represent individual `tiles'. The CFHTLenS fields W1, W2, W3, and W4 are covered with the maximum number of non-overlapping tiles that fit within their respective areas. Given the area of the four CFHTLenS mosaic fields, one can only cover W1 and W3 with $4$ tiles each, and W2 and W4 with one tile each. The final simulated area is therefore $124$ deg$^2$; which is similar to the $154$ deg$^2$ covered by CFHTLenS. Note that since each of the tiles used in our simulations is an independent line of sight, we expect sampling variance to affect the mock catalogues less than the four CFHTLenS fields, where only four fields are statistically independent.

\item The 2D positions on the sky of all galaxies in CFHTLenS are preserved. The orientation of each galaxy's ellipticity is randomized, but its amplitude is held fixed as this is used as the only source of shape noise in the mock catalogue. The redshift of each galaxy is resampled using redshift probability distribution obtained from the photometric redshift distribution function. We refer the reader to \cite{2012arXiv1212.3327B} for the details. The galaxies are then placed on the mosaic tiles and each galaxy's redshift determines uniquely the simulated shear and convergence. This is calculated using Eqs.(\ref{gammadef}) and (\ref{kappadef}) with the reduced shear computed in Eq.(\ref{gdef}). The final `observed' ellipticity in the mock catalogue is obtained by Eq.(\ref{gdef2}). The masks are applied to the mock catalogue (i.e. no `missing galaxy' has been added to fill in the gaps), as are all the other characteristics of the survey (e.g. the weight associated with galaxy shape measurements is also preserved).

\item Following the procedure described in Section 2, the mass reconstruction is performed on a regular $512\times 512$ grid, which is a $2\times 2$ re-binning of the simulation's native pixel grid. The mass reconstruction is performed for $5$ different smoothing scales, with a radius ranging from approximately $2$ to $9$ arcminutes.

\item For some of the following tests performed on simulations, we will be using the noise-free convergence map, obtained from the stacking of convergence maps at difference source redshifts using the redshift distribution from (ii) above. These noise-free maps are obtained directly from the simulated light cone, prior to the construction of the mock catalogues. For this reason, and to make a clear distinction with the quantity $\kappa_E$, we call the stacked noise-free convergence map $\kappa_{\rm sim}$.

\end{enumerate}

Figure \ref{simul1} shows an example of a mass reconstruction of one the tiles.
The background image with the white contours shows a noisy mass reconstruction; the white contours represent the $1,2,3, 4\sigma$ levels, where $\sigma$ is the convergence rms determined from the map. The black contours represent the $1,2,3, 4\sigma$ levels on the noise-free convergence map $\kappa_{\rm sim}$. This figure illustrates qualitatively how realistic shot noise (ellipticity noise) noticeably affects the position, amplitude, and sometimes even the presence or absence of peaks in the reconstructed map. Many reconstructed peaks do not match a real mass peak and the converse is also true. Although a quantitative analysis of peaks is left for another paper, this illustration is consistent with earlier work \citep{2000MNRAS.313..524V,2002A&A...384..743A} showing that individual peaks are relatively noisy objects, with a higher chance of being a coincidence ($20$ per cent for a $3 \sigma$ peak) than what Gaussian statistics would predict in the field. Masks are also shown in Figure \ref{simul1} as the black areas with sharp boundaries, which shows that our mass reconstruction procedure does not generate catastrophic edge effects near the masks. The rest of this Section explores the reliability of the mass reconstruction quantitatively.

\subsection{Analysis of the mock catalogues}

\begin{figure}
 \begin{center}
   \includegraphics[angle=-90,width = 8.6cm]{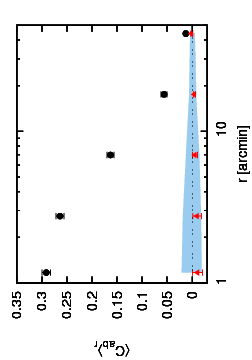}
   \caption{For a smoothing scale of $\theta_0=2.5$ arcmin, the different sets of points show the cross-correlation profile $\langle{\cal C}_{a; b}\rangle_{\rm r}$ between two convergence maps $\kappa_a$ and $\kappa_b$, where $a$ and $b$ are one of "obs", "$B$", "sim" or "ran". The black filled circles show $\langle{\cal C}_{{\rm obs}; {\rm sim}}\rangle_{\rm r}$, the red filled triangles show $\langle{\cal C}_{{\rm obs}; B}\rangle_{\rm r}$. Error bars for the filled circles and triangles are the $1\sigma$ rms of the average over the $10$ lines of sight. The light blue filled area shows the $1\sigma$ region of $\langle{\cal C}_{{\rm obs}; {\rm ran}}\rangle_{\rm r}$ averaged over $100$ random noise realizations. The $1\sigma$ region in the latter also represents the deviation of the average.}
\label{crossc}
\end{center}
\end{figure}

In this Section, we use the simulations and their reconstructed lensing maps to verify that the level of shape noise and masks in the CFHTLenS data will not introduce systematic errors in the statistical properties of the reconstructed convergence map. This will be performed by running two distinct tests.

For the first test, we focus on the cross-correlation between two maps $\kappa_a$ and $\kappa_b$, where $a$ and $b$ can be any of the signal $\kappa_{\rm obs}$, noise-free $\kappa_{\rm sim}$, galaxy rotated $\kappa_\perp$, or randomized $\kappa_{\rm ran}$ maps. The cross-correlation map ${\cal C}_{a; b}$ is given by:

\begin{equation}
{\cal C}_{a; b}={\kappa_a \star \kappa_b\over \sqrt{\langle \kappa_a^2\rangle_0}\sqrt{\langle \kappa_b^2 \rangle_0}},
\label{crosscorrel}
\end{equation}
where $\langle...\rangle_0$ denotes the zero lag value of the auto-correlation map. This definition guarantees that ${\cal C}_{a;b}$ is normalized, i.e. the central pixel of ${\cal C}_{a;b}$ is equal to one if $\kappa_a=\kappa_b$. ${\cal C}_{a;b}$ is then azimuthally averaged within annuli as a function of the distance $\rm r$ from the central pixel; this quantity is called $\langle{\cal C}_{a;b}\rangle_{\rm r}$.

Figure \ref{crossc} shows $\langle{\cal C}_{a;b}\rangle_{\rm r}$ for different combinations of reconstructed and noise-free maps for the Gaussian smoothing scale $\theta_0=2.5$ arcmin. The filled triangles show $\langle{\cal C}_{E; \perp}\rangle_{\rm r}$, the cross-correlation profile between the reconstructed mass map and the galaxy rotated map, averaged over the $10$ lines of sight of the mock catalogue (see Section 3.2). The error bars, which represent the error on the average profile, show that $\langle{\cal C}_{{\rm obs}; \perp}\rangle_{\rm r}$ is consistent with zero. The solid area shows the scatter of $\langle{\cal C}_{\rm obs; ran}\rangle_{\rm r}$ averaged over $100$ random realizations of $\kappa_{\rm ran}$. This quantity is also consistent with zero, which confirms that the field boundary, edges, and masks do not generate systematic effects, even for pure noise reconstructions. The filled circles show the cross-correlation profile between the mass reconstruction and the noise-free map $\langle{\cal C}_{\rm obs; sim}\rangle_{\rm r}$. The error bars illustrate the dispersion of the average over the $10$ lines of sight; the zero-lag cross-correlation coefficient is only $0.3$, which is another indication that mass maps are noisy, especially for small smoothing scales. Larger smoothing scales, not shown here, lead to a stronger cross-correlation amplitude, but also a larger correlation length.
This first test demonstrates the absence of systematic effects (e.g. spurious peaks) around masks and low-density areas; if these were present, we would expect a significant non-zero residual cross-correlation between the reconstructed mass map $\kappa_{\rm obs}$ and $\kappa_\perp$ or the pure noise reconstruction.

\begin{figure*}
 \begin{center}
   \includegraphics[angle=-90,width = 17cm]{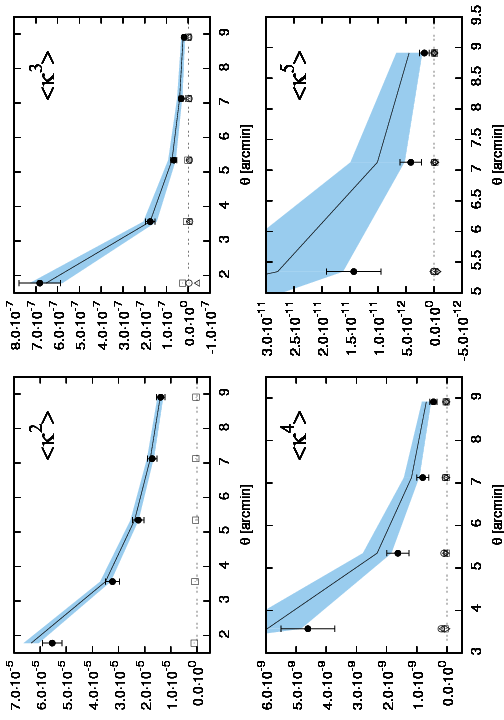}
   \caption{Moments of the convergence $\langle\kappa^n\rangle$, $n=2,3,4,5$, measured on the simulations. Filled circles show the de-noised moments for the reconstructed mass map $\kappa_{\rm obs}$. The error bars show the $1\sigma$ deviation of the mean of the $10$ lines of sight. The solid line inside the light blue region shows the noise-free moments measured on $\kappa_{\rm sim}$, and the filled light blue area shows the $1\sigma$ deviation of the mean of the $10$ noise-free maps. Open symbols show all the possible de-noised combinations of $\kappa_{\rm obs}$ and $\kappa_\perp$: $\langle\kappa_B^2\rangle_{\theta_0}$ for the top left-panel, $\langle\kappa_B^3\rangle_{\theta_0}$, $\langle\kappa_E\kappa_B^2\rangle_{\theta_0}$, $\langle\kappa_E^2\kappa_B\rangle_{\theta_0}$ for the top-right panel, $\langle\kappa_B^4\rangle_{\theta_0}$, $\langle\kappa_E\kappa_B^3\rangle_{\theta_0}$, $\langle\kappa_E^2\kappa_B^2\rangle_{\theta_0}$, $\langle\kappa_E^3\kappa_B\rangle_{\theta_0}$ for the bottom-left panel and $\langle\kappa_B^5\rangle_{\theta_0}$, $\langle\kappa_E\kappa_B^4\rangle_{\theta_0}$, $\langle\kappa_E^2\kappa_B^3\rangle_{\theta_0}$, $\langle\kappa_E^3\kappa_B^2\rangle_{\theta_0}$, $\langle\kappa_E^4\kappa_B\rangle_{\theta_0}$ for the bottom-right panel.}
\label{simul2}
\end{center}
\end{figure*}

The second test consists of looking at various higher-order moments of the convergence. Like the traditional measurement of the second order moment of the shear $\langle \gamma^2\rangle$ (here $\gamma^2=\gamma_1^2+\gamma_2^2$), we want to measure the true moments of the convergence $\langle \kappa_E^n\rangle$, with $n=2,3,4,5$. A measurement of the convergence moments from noisy reconstructed mass maps has never previously been reported. However, there is a complication with the measurement of convergence moments: the convergence can only be measured within a smoothing window, unlike the shear, which is given for each individual galaxy. Therefore, the convergence noise is correlated for different points on the grid, with the result that the moments of $\kappa_{\rm obs}$ and $\kappa_E$ are not equal.
A shot noise removal, or {\it de-noising}, is therefore required in order to remove this bias from the observed convergence moments $\langle \kappa_{\rm obs}^n\rangle$.

We choose a direct approach for the de-noising procedure, which consists of removing the correlated noise contribution from the moments measured on $\kappa_{\rm obs}$. Fortunately, two independent residual systematics tests can be performed; one on the `mass map' of the rotated galaxies $\kappa_\perp$, and the other by comparing the de-noised moments of $\kappa_E$ to the moments of the noise-free maps $\kappa_{\rm sim}$. Figure \ref{crossc} shows the absence of cross-correlation between $\kappa_{\rm obs}$ and the noise $\kappa_{\rm ran}$ which means that, to first approximation, the signal and noise can be treated as statistically independent \citep{2000MNRAS.313..524V}. We use this lack of correlation to derive relatively straightforward relations between the moments of $\kappa_{\rm obs}$ and $\kappa_E$. For instance, the observed second order moment $\langle\kappa^2_{\rm obs}\rangle$ is the quadratic sum of $\langle\kappa^2_E\rangle$ and $\langle\kappa^2_{\rm ran}\rangle$. The extension of this relation to higher order moments defines our de-noising procedure. Moments of the noise maps $\kappa_{\rm ran}$, needed for this step, are measured from a large number of pure noise reconstructions. The relations below show how, for each smoothing scale $\theta_0$, the observed moments of $\kappa_{\rm obs}$ are related to the true moments of $\kappa_E$ and the moments of the noise map $\kappa_{\rm ran}$:

\begin{eqnarray}
&&\langle \kappa_{\rm E}^2\rangle_{\theta_0}=\left(\kappa^2_{\rm obs}\right)_{\theta_0}-\overline{\left(\kappa_{\rm ran}^2\right)}_{\theta_0} \cr
&&\langle \kappa_{\rm E}^3\rangle_{\theta_0}= \left(\kappa^3_{\rm obs}\right)_{\theta_0} \cr
&&\langle \kappa_{\rm E}^4\rangle_{\theta_0}= \left(\kappa^4_{\rm obs}\right)_{\theta_0}
-6\overline{\left(\kappa^2_{\rm obs} \kappa_{\rm ran}^2\right)}_{\theta_0}-\overline{\left(\kappa_{\rm ran}^4\right)}_{\theta_0} \cr
&&\langle \kappa_{\rm E}^5\rangle_{\theta_0}= \left(\kappa^5_{\rm obs}\right)_{\theta_0}-10\overline{\left(\kappa^3_{\rm obs}\kappa_{\rm ran}^2\right)}_{\theta_0}.
\end{eqnarray}
For clarity, $\left(...\right)_{\theta_0}$ denotes the moment average over {\it one} map, $\overline{\left(...\right)}_{\theta_0}$ is averaged over several noise maps $\kappa_{\rm ran}$, and $\langle...\rangle_{\theta_0}$ is the de-noised moment.
We limit our analysis to higher order moments up to the fifth order. Beyond the fifth order, we find that the measurements become too noisy with the CFHTLenS data.
The Appendix includes the expressions of the residual systematics moments, i.e. all the combinations of $E$ and $B$ modes that we expect to vanish if residual systematics are negligible for $n=2,3,4,5$. As the measurements will show, the residual systematics are indeed consistent with zero.
Our de-noising approach automatically takes care of varying noise across the field because the noise rms for different positions on a map is preserved for the different noise realizations. This is due to the fact that we keep the galaxy's position and shape noise unchanged.
A further improvement of the de-noising technique was implemented by taking into account the fraction of pixels not masked within each smoothing window. We achieve this by constructing a filling factor map computed from the smoothed grid with pixels set to one, and then masking and smoothing with the same window that was applied to the lensing data for the mass reconstruction. Using this filling factor map, regions around masks in the lensing reconstructed maps are down-weighted without biasing the signal. In order to eliminate the noisiest regions, we apply a cut-off of $50$ per cent such that pixels below this threshold are excluded from the moments analysis. We verified that the application of this cut-off has a negligible effect on the measurement with a cut-off varying from $0$ to $80$ per cent. Of course, high cut-off values boost the noise and it becomes difficult to evaluate changes in the signal. This procedure guarantees that pixels in the mass map contribute proportionally to the unmasked area located in a smoothing window. We will compare the de-noised moments of $\kappa_{\rm obs}$ with the noise-free moments measured on $\kappa_{\rm sim}$.

Figure \ref{simul2} shows the convergence moments as measured from the mock catalogues using the de-noising technique described earlier. We see that the de-noised moments $\langle \kappa_E^n\rangle$ (filled circles) are consistent with the noise-free moments $\langle\kappa_{\rm sim}^n\rangle$. Note that the noise-free moments represent the true answer in the sense that they are directly computed from the N-body simulations, without mass reconstruction, and they therefore contain no noise, no discrete sampling and no mask. On the other hand, the moments $\langle \kappa_E^n\rangle$ are de-noised moments obtained from the noisy mass reconstruction maps that include masks. The two moments agree, which demonstrates that our mass reconstruction procedure leads to reliable mass maps that preserve the statistical and cosmological information. In Figure \ref{simul2}, the de-noised moments' errors were computed from the variance between the $10$ lines of sight and divided by $\sqrt{10}$. The noise-free moments' errors are shown by the filled regions around the solid lines, and are also divided by $\sqrt{10}$. The open symbols show, for the different moments, the residual systematics obtained from all the possible combinations of the reconstructed maps $\kappa_{\rm obs}$ and $\kappa_\perp$, where the de-noised moments were computed using the expressions derived in the Appendix. All moments involving one or more of the rotated galaxies maps $\kappa_\perp$ are consistent with zero, showing that the $B$-mode is consistent with pure noise once it has been de-noised.

These results from mock catalogues validate our approach for the CFHTLenS data and demonstrate that for the $154$ deg$^2$ of the survey, our reconstruction process is stable. Furthermore, we have shown that realistic masking geometry does not alter the reconstruction, and therefore we are able to reliably quantify some of the most basic statistics of the projected mass density.


\subsection{Beyond the moments: Convergence PDF and peak statistics}

The first few convergence moments give only a partial description of the histogram of the convergence, which is also called the $1$-point Probability Distribution Function (PDF). However, the $1$-point PDF, along with its extensions to higher order, contains additional information about the moment hierarchy, and therefore about the structure formation process. For instance, \citet{1995ApJ...442...39J} demonstrated that a combination of different moments of the PDF, through the Edgeworth expansion, probes different aspects of the gravitational collapse. General characteristics of the PDF, such as the height and the minimum $\kappa$ of the convergence histogram, are important features that are not easily captured by convergence moments. Therefore, it would be interesting to use the convergence PDF itself as a cosmological probe. This possibility has been theoretically explored \citep{2004MNRAS.354.1146V,2004MNRAS.350...77M} using the aperture filtered shear, the same filter which transforms the shear $\gamma$ into a local, scalar quantity \citep{1996MNRAS.283..837S}. Doing this analysis on the convergence $\kappa$ PDF, instead of the aperture filtered shear, would exploit the fact that the long wavelength modes are preserved; this analysis with the CFHTLenS mass maps is left for a forthcoming study.

\begin{figure}
 \begin{center}
   \includegraphics[angle=-90,width = 8.5cm]{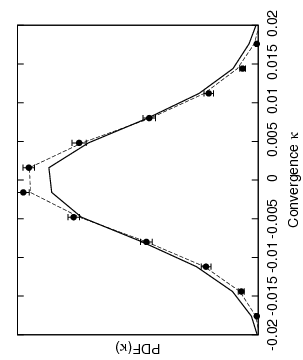}
   \caption{Probability distribution function histogram of the convergence for the reconstructed map $\kappa_{\rm obs}$ (solid line) and for the galaxy rotated map $\kappa_\perp$ (dashed line). The filled circles with error bars are obtained from an average of $10$ noise realisation maps $\kappa_{\rm ran}$ where the galaxy orientations have been randomised before the mass reconstruction.}
\label{simul3}
\end{center}
\end{figure}

Nevertheless, we can already illustrate with our simulations the expected level of $1$-point PDF signal-to-noise. Figure \ref{simul3} shows the measured $1$-point PDF for $\kappa$ compared to the $\kappa_\perp$ $1$-point PDF. The average $1$-point PDF obtained from pure noise reconstructions (with error bars) is also shown, and the $B$-mode PDF is consistent with pure noise. This result is in agreement with Figure \ref{simul2}, which shows a negligible $B$-mode for the convergence moments. One can also clearly see that the cosmological signal broadens the PDF compared to pure noise (or $B$-mode) PDFs. This $1$-point PDF study opens up a new statistical analysis opportunity for gravitational lensing surveys, which remains to be exploited. Alternative probes for cosmology using the convergence PDF could be constructed, such as using the minimum value of the convergence found in voids (modulo the convolution by the noise) as a direct measurement of the mass density parameter $\Omega_0$, which is relatively insensitive to the non-linear clustering.

Another interesting use of convergence maps is the study of peak statistics. The idea of using peaks in dark matter maps to constrain the halo mass function was first proposed in \cite{2000MNRAS.313..524V,2000ApJ...530L...1J}. This analysis relies heavily on the properties of peak statistics in Gaussian noise developed for the cosmic microwave background \citep{1987MNRAS.226..655B}. The peak statistic theory has recently been pursued further \citep{2010A&A...519A..23M,2012MNRAS.423.1711M}. Peak statistics provides a higher level of statistical analysis of maps, where we expect the sensitivity to cosmology and residual systematics to be different from the statistical analysis using moments of the shear or convergence. It is important to distinguish between the maps constructed in this paper and the maps being discussed in \cite{2012MNRAS.423.1711M}, which use the aperture mass filter, originally defined in \cite{1996MNRAS.283..837S}. The aperture mass filter is a pass-band filter, as opposed to the Gaussian (or top-hat) smoothing window used in our study, which is a low-pass filter. The low-pass filter preserves the large scale modes, hence keeping visible the voids and large overdensities, unlike what happens with a pass-band filter (see Section 4.4 below). The peak statistics of a low-pass map allows the study of the dark matter halo mass function as a function of the larger scale dark matter environment, which is essentially not possible with a band-pass filter. The two approaches are complementary and it would be beneficial in the future to unify them into a single approach to making mass maps. Peak statistics, convergence PDF, and the morphological analysis of large scale structure from the CFHTLenS data are left for future studies.

\section{Results from CFHTL\lowercase{en}S}

Residual systematics in the shear signal, based on the selected fields that passed the systematics tests (the "good" fields) are given in \cite{2012MNRAS.427..146H}. In total, $129$ MegaCam pointings (out of $171$) passed the residual systematics tests. The convergence moments presented in this paper are based on these good fields, while the mass reconstruction is performed on all the fields. Each galaxy in the CFHTLenS catalogue has a shear estimate $\eg^{\rm obs}$ and a weight $w$ \citep{2013MNRAS.429.2858M}, a calibration factor $m$ \citep{2012MNRAS.427..146H}, and a photometric-redshift probability distribution function \citep{2012MNRAS.421.2355H}. In the following sections we describe how these elements are integrated into the mapmaking process.

For the lensing mass reconstruction, source galaxies are selected in the redshift range $z=[0.4,1.1]$, which is sufficiently broad to avoid the complication of intrinsic alignment \citep{Heymans2013}, but narrow enough to guarantee that the 3-point signal is detectable at a few sigma \citep{Vafaei10}.

\subsection{Mass reconstruction}

Mass reconstruction on the CFHTLenS data follows the same procedure as described in Section 2 using the KS93 algorithm \citep{1993ApJ...404..441K}. The regular grid over which the mass reconstruction and smoothing are performed consists of square pixels with an area of approximately $1$ arcmin$^2$.
The average of the galaxy ellipticities is first calculated within each pixel, using the calibration correction implemented in \cite{2013MNRAS.429.2858M} in order to account for the constant correction $c_2$ \citep{2012MNRAS.427..146H}, the multiplicative shear measurement bias $(1+m_i)$, and the weighting $w_i$ for each galaxy located at $\thetavc_i$ within that pixel. The shear estimate per galaxy is not divided by $(1+m_i)$, but instead the average $\bar \eg_{\rm pix}$ in each pixel is given by the weighted sum over the galaxies in that pixel:

\begin{equation}
\bar\eg_{\rm pix}={\sum_{i=1,2} w_i\eg_i(\thetavc_i)\over\sum_{i=1,2} w_i (1+m_i)} .
\end{equation}
The smoothing of this grid-averaged ellipticity map is then performed using the Gaussian window function given by Eq.(\ref{Wgauss}). Mass maps are reconstructed with the following smoothing scales: $1.8$, $3.5$, $5.3$, $7.1$, and $8.9$ arcmin. The final maps have a size of $512\times 449$, $512\times 505$, $512\times 495$, and $512\times 502$ for W1, W2, W3, and W4 respectively, which is chosen to be close to the grid size that was used on the simulations. The fact that the angular resolution is different for the different fields has no effect on our results.
Mass reconstruction is also performed with the galaxies rotated by $45$ degrees in order to probe the $B$-modes. As a sanity check, we computed the cross-correlation map between $\kappa_{\rm obs}$ and $\kappa_\perp$, following Eq.(\ref{crosscorrel}), which shows that the two maps are uncorrelated, hence supporting the conclusion that the $B$-mode in the data is consistent with zero.

\subsection{Cosmic statistics on CFHTL\lowercase{en}S mass maps}

\begin{figure}
 \begin{center}
   \includegraphics[angle=-90,width = 8.6cm]{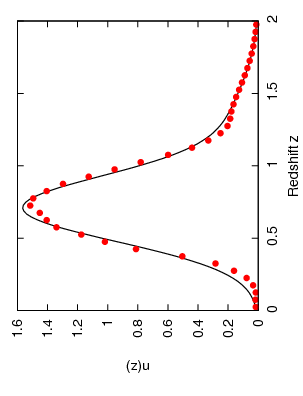}
   \caption{Redshift distribution of the galaxies from the CFHTLenS catalogue that are used for the measurement of the convergence moments. Data points show the actual probability distribution obtained from the photometric redshifts and the solid line shows the best-fitting function defined in Eq.(\ref{nofzmodel}). The best-fitting parameter values are $a=1.50$, $b=0.32$, $c=0.20$, and $d=0.46$.}
\label{nofz}
\end{center}
\end{figure}

Following the procedure described in Section 3.2, $100$ independent pure noise maps $\kappa_{\rm ran}$ are reconstructed for each CFHTLenS field and each smoothing scale. The pure noise maps are necessary for the de-noising of the convergence moments. This paper is the first to measure cosmic shear statistics directly from reconstructed mass maps and it is therefore important to quantify the reliability of the measurements against predictions. Theoretical predictions for the second and third order moments are calculated using Eqs.(\ref{kappa2def}) and (\ref{kappa3def}). The key ingredient for these predictions is an accurate redshift distribution $n_S(z)$ of the galaxies. In order to estimate $n_S(z)$, we use the probability distribution function of the photometric redshifts \citep{2012MNRAS.421.2355H}, the robustness of which has been thoroughly tested in \citet{2012arXiv1212.3327B}. The implementation of this approach in other CFHTLenS papers, where the second order moment of the shear has been used to constrain cosmological parameters, shows that the redshift distribution derived from CFHTLenS photometric redshifts provides a robust and consistent interpretation of the weak lensing data as a function of source redshift \citep{2012arXiv1212.3327B,IIGIGG,2013MNRAS.tmp..735K,2013MNRAS.429.2249S}.

The data points in Figure \ref{nofz} show the stacked redshift distribution PDF for our galaxy selection with photometric redshifts $0.4<z<1.1$, the same galaxy selection used for the lensing mass reconstruction. In order to make predictions for the lensing statistics, we fit the observed redshift distribution with the following four-parameter double-Gaussian model:

\begin{equation}
n_S(z)=a\times\exp\left(-{(z-0.7)^2\over b^2}\right)+c\times\exp\left(-{(z-1.2)^2\over d^2}\right).
\label{nofzmodel}
\end{equation}
Figure \ref{nofz} shows that this is a reasonable fit to the data, in particular in capturing the long, slowly-decreasing tail at high redshifts. The best fit values are $a=1.50$, $b=0.32$, $c=0.20$, and $d=0.46$.

\begin{figure*}
 \begin{center}
   \includegraphics[angle=-90,width = 17cm]{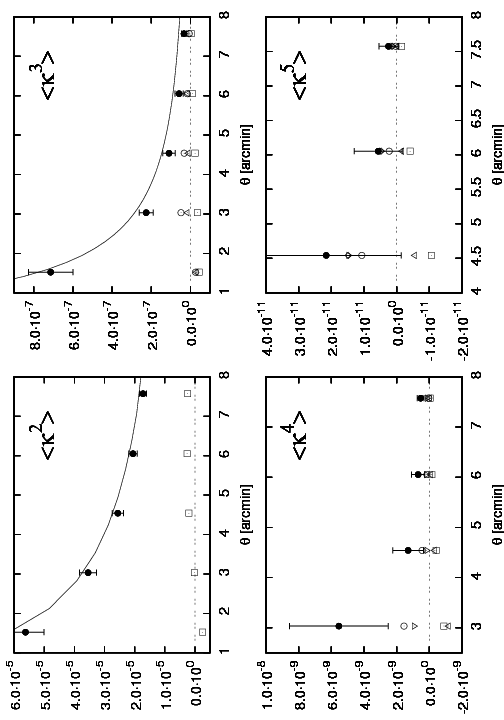}
   \caption{Moments of the convergence $\langle\kappa^n\rangle$, $n=2,3,4,5$ measured on the CFHTLenS data. Error bars show the $1\sigma$ deviation from the mean of the the four CFHTLenS fields. Solid lines are the moments measured from the signal maps and de-noised using the procedure described in Section 3.2. Open symbols show the different de-noised combinations of the signal map $\kappa_{\rm obs}$ and systematics map $\kappa_{\perp}$, similar to that shown in Figure \ref{simul2} for the simulations. The solid line shows the second order moment (top-left) and third order moment (top-right) predictions from Eqs.(\ref{kappa2def}) and (\ref{kappa3def}) using the WMAP7 cosmology (see text in Section 4.3).}
\label{cfhtlensstat}
\end{center}
\end{figure*}

Figure \ref{cfhtlensstat} shows the second to fifth order moments of the convergence measured on CFHTLenS lensing data. The solid lines show the predictions for the second and third order moments using the redshift distribution derived above. The predictions are not a {\it fit} to the data. They are computed using the CAMB non-linear power spectrum\footnote{http://camb.info/} for the $\Lambda {\rm CDM}$ model $\Omega_b=0.0471$, $\Omega_{\rm CDM}=0.2409$, and $\Omega_\Lambda=0.712$ for the baryonic, dark matter, and cosmological constant density parameters respectively. The matter power spectrum is normalised to $\sigma_8=0.792$ at $z=0$. The other cosmological parameters, such as the power spectrum tilt, running spectral index, etc., are identical to the best-fitting values from WMAP7 \citep{2011ApJS..192...18K}. This particular choice of parameters is consistent with the parameter values obtained in the CFHTLenS companion papers \citep{2012arXiv1212.3327B,IIGIGG,2013MNRAS.tmp..735K,2013MNRAS.429.2249S,Kitching2013}. The predictions agree well with measurements for the second and third order statistics, which gives us a high level of confidence in the reliability of mass maps to extract cosmological information and confirms the analysis performed on the shear. The error bars represent the sampling variance between the four CFHTLenS fields, accounting for the different image sizes; therefore, they capture the total error budget (including statistical variance). Since there are no reliable fully non-linear predictions for the fourth and fifth order moments, a full cosmological analysis including all moments must rely on ray-tracing simulations. This work is left for a future study in which the properties of the full convergence probability distribution function will be investigated. As Figure \ref{cfhtlensstat} illustrates, the CFHTLenS data show a marginal detection of the fourth order moment and no detection of the fifth. The residual systematics are consistent with zero for all moments.


\subsection{The connection between large scale dark matter and baryons}

\subsubsection{Construction of the mass map predicted from baryons}

Sections 3 and 4 describe the mass reconstruction process in the presence of realistic noise and masks. A cosmological signal is clearly measured on the CFHTLenS data and the level of residual systematics is shown to be consistent with zero. In this section, we are interested in the comparison between the reconstructed dark matter and the matter distribution of the stellar content. The connection between dark matter and baryons is quantified in the CFHTLenS companion papers \citep{2013arXiv1301.7421G,velander2013} at the galaxy and galaxy-group scales. Here we are interested in the connection between dark matter and baryons at much larger scales. For example, we would like to explore whether there are large scale features common to both the dark matter and the baryons, such as voids. To this end we develop a new approach, which consists of first predicting the dark matter map {\it from} the stellar mass distribution \citep{2001ApJ...556..601W}, and then comparing the peak distribution from the predicted $\kappa$ map to the lensing map.

In order to construct a predicted $\kappa$ map, a Navarro-Frenk-White dark matter density profile \citep{1997ApJ...490..493N} is assigned to each galaxy in the CFHTLenS catalogue; the procedure is detailed below. There is no need to separate the galaxy population into sources and lenses because most galaxies act as both due to our broad redshift selection. Galaxies are assigned a dark matter mass following the relation between stellar mass and halo mass from \citet{2012ApJ...744..159L}, which provides a relation for redshifts between $0.2$ and $1$, covering almost the entire redshift range of galaxies in our work. The only complication is that this relation is only provided for central galaxies, while in our sample a large fraction are satellites. Unfortunately, it is impossible to separate the central galaxies from the satellites on an individual basis; it can only be done statistically \citep{velander2013}. We therefore proceed by assuming that all galaxies are central galaxies and all follow the stellar mass to halo mass relation in \citet{2012ApJ...744..159L}. This will overestimate our predicted total mass, but to a first approximation it should not dramatically affect the relative distribution of mass. Based on Figure 4 from \citet{2011ApJ...738...45L}, we anticipate that, on average, there are roughly one to two satellite galaxies for every central galaxy. This should lead us to overestimate the total mass by roughly a factor of $2$ to $3$. The exact calculation is not needed as we are only interested in an order of magnitude estimate of how wrong our predicted convergence can be. In a future work, the same strategy will be applied to clusters instead of individual galaxies, which should mitigate this effect. In order to complete our convergence prediction from the galaxies, we need to assign a concentration to each halo. To this end, we use the mass-concentration relation calibrated from numerical simulations in \cite{2011MNRAS.411..584M}.

\begin{figure}
 \begin{center}
   \includegraphics[angle=-90,width = 8.5cm]{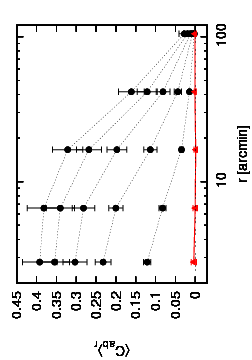}
   \caption{The five black dotted lines with points and error bars show the cross-correlation function $\langle {\cal C}_{\rm obs; gal}\rangle$ with CFHTLenS data. The $\kappa_{\rm obs}$ and $\kappa_{\rm gal}$ maps have been smoothed with a Gaussian window of $1.8$, $3.5$, $5.3$, $7.1$, and $8.9$ arcmin from bottom to top. The data points are the mean over the four CFHTLenS fields and the error bars represent the error on the mean. The red line with triangle data points show $\langle {\cal C}_{\perp \rm ; obs}\rangle$ for a smoothing scale of $1.8$ arcmin. The other smoothing scales are also consistent with zero, but are not shown here for the sake of clarity.}
\label{crosscorreldata}
\end{center}
\end{figure}

\begin{figure*}
 \begin{center}
   \includegraphics[width=\textwidth]{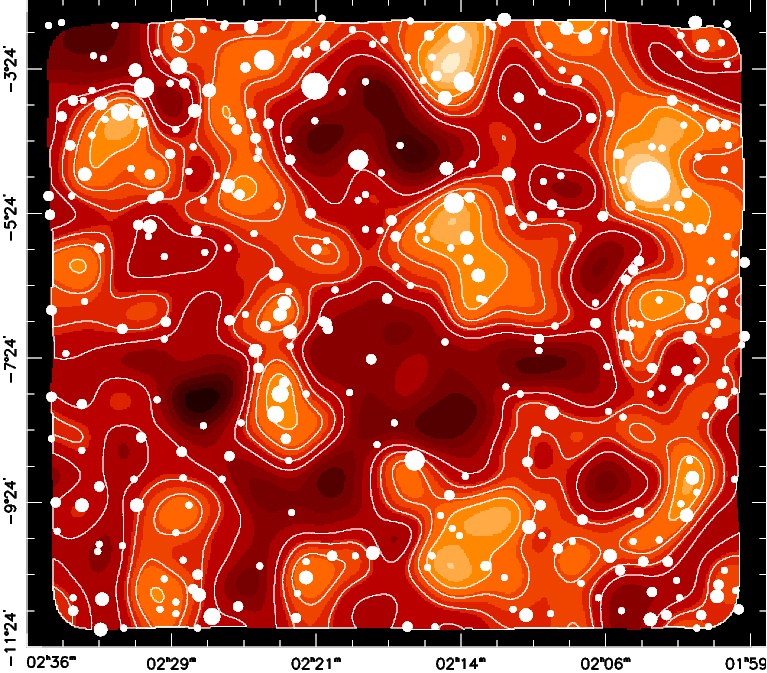}
   \caption{ Mass maps for the  W1 field. The continuous map with contours shows the mass reconstructed from gravitational lensing. Contours indicate the $1, 2, 3$, and $4$ $\sigma$ on this map, where $\sigma$ is the rms of the convergence. Open circles indicate the position of peaks in the predicted mass map, constructed from galaxies as described in Section 4.3. The circle size is proportional to the peak height. The field of view is approximately $9\times 8$ deg$^2$.}
\label{W1peaks}
\end{center}
\end{figure*}

At this stage, each galaxy in the CFHTLenS catalogue is associated with a dark matter halo of known concentration and mass. The last step is to apply the lensing kernel in order to predict the convergence based on the redshifts of the lenses and sources. Every galaxy is simultaneously both a lens and a source, depending on whether it is in the background or the foreground relative to other galaxies. One can then compute for each galaxy a predicted convergence based on the foreground mass distribution coming from all galaxies located at lower redshift. For a source galaxy at location $\thetavc$ on the sky with redshift $z_S$ and $N$ foreground lenses at redshifts $z_{L_i}$ the total convergence predicted from the baryonic distribution is given by:

\begin{equation}
\kappa_{\rm gal}(\thetavc)=\sum_{i=1}^N {\Sigma_i(\left|\thetavc-\thetavc_i\right|)\over \Sigma_{crit}(z_{L_i},z_S)}-\bar\kappa_{\rm gal},
\label{kappabar}
\end{equation}
where $\Sigma_i(\left|\thetavc-\thetavc_i\right|)$ is the projected halo mass of lens $i$ centred at $\thetavc_i$, and $\Sigma_{crit}(z_{L_i},z_S)$ is the critical density given by:

\begin{equation}
\Sigma_{crit}(z_{L_i},z_S)={c^2\over 4\pi G}{f_K(w_S)\over f_K(w_L) f_K(w_S-w_{L_i})}.
\end{equation}
Note that the average predicted convergence $\bar\kappa_{\rm gal}$ is calculated only after all haloes have been assigned to the galaxies.
The critical density depends on the observer-lens, lens-source, and observer-source angular diameter distances $f_K(w_L)$, $f_K(w_S-w_{L_i})$, and $f_K(w_S)$ respectively. The sky-average predicted convergence is set to zero by subtracting the mean $\bar\kappa_{\rm gal}$ in Eq.(\ref{kappabar}). We assume that the weak lensing approximation applies, which means that the convergence from the different lenses can be added linearly. It is important to emphasize that the lens redshift goes down to the lowest reliable value $z=0.2$, and that the sources only cover redshifts $z=0.4$ and higher (in order to be consistent with our source galaxy selection outlined in Section 4.2).

\begin{figure*}
 \begin{center}
   \includegraphics[angle=-90,width = 8.7cm]{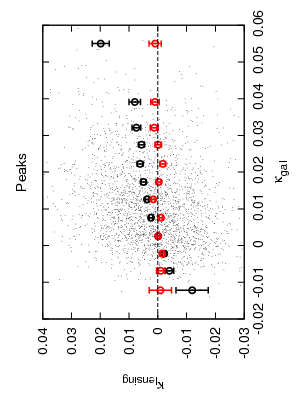}
      \includegraphics[angle=-90,width = 8.7cm]{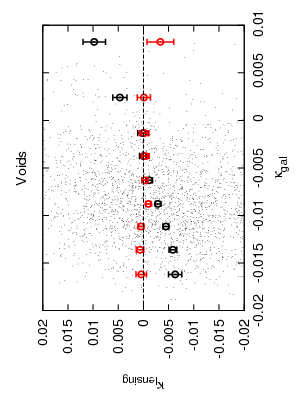}
   \caption{Left panel: The x-axis values show, for all W fields, the predicted convergence $\kappa_{\rm gal}$, taken at peak locations. The y-axis shows the lensing $\kappa_{\rm lensing}$ values measured from either the mass reconstruction map $\kappa_{\rm obs}$ or the rotated galaxies reconstructed map $\kappa_\perp$. The lensing $\kappa_{\rm lensing}$ values are taken at positions given by the peaks location in the $\kappa_{\rm gal}$ map. Individual dots on the figure represent individual peaks, and the black open circles show the binned average. The red open circles show the binned average of the $\kappa_{\rm gal}$ peaks values versus $\kappa_\perp$. Error bars always represent the dispersion on the mean. Right panel: The x-axis now shows the predicted convergence $\kappa_{\rm gal}$ values taken at the troughs locations. The y-axis shows the lensing convergence values taken at the $\kappa_{\rm gal}$ troughs locations from either $\kappa_{\rm obs}$ or $\kappa_\perp$. The detection of voids is manifest, trough the very significant negative values of $\kappa_{\rm obs}$ correlated with the $\kappa_{\rm gal}$ troughs.}
\label{baryonspeaks}
\end{center}
\end{figure*}

The convergence predicted from the baryonic content $\kappa_{\rm gal}(\theta)$ is assigned to each galaxy in the CFHTLenS catalogue. Following Section 4.2, the $\kappa_{\rm gal}(\theta)$ are placed on the same regular grid that is used for the lensing mass reconstruction. The {\em lens}fit weighting is used to determine the average $\bar\kappa_{\rm gal}(\theta)$ within each pixel, and the same Gaussian smoothing is applied. For most galaxies, the size $r_{200}$ of a single galactic halo is comparable to the size of a pixel, which is of the order of half an arcminute, in the final resolution map. The galaxies in the masked regions therefore do not impact on our analysis as the extension of their halo is small \footnote{Note that the haloes from low redshift galaxies do extend over several pixels, but their lensing efficiency is small due to their proximity to the observer}. The Gaussian smoothing window is sometimes truncated by the masks, but this effect is minimal because of the filling factor cut of $50$ per cent applied to the predicted mass map. Furthermore, this is a random, zero net effect, leading only to larger noise around the image masks, but not to a bias in the projected convergence.

\subsubsection{Comparing the lensing with the predicted dark matter maps}

In this Section, we perform a comparison of the $\kappa_{\rm obs}$ and $\kappa_{\rm gal}$ maps. Following an approach similar to Section 3.3, when comparing noise-free and noisy simulated data, we first compute the cross-correlation profile between the lensing reconstructed mass map $\kappa_{\rm obs}$ and the predicted map $\kappa_{\rm gal}$ and between the rotated galaxies map $\kappa_\perp$ and $\kappa_{\rm gal}$. These profiles are called $\langle {\cal C}_{\rm obs; gal}\rangle$ and $\langle {\cal C}_{\perp \rm ; obs}\rangle$ respectively, as defined by Eq.(\ref{crosscorrel}). Figure \ref{crosscorreldata} shows that the correlation increases for larger smoothing windows, and that the overall correlation level always remains below $50\%$. The relatively low cross-correlation coefficient for small smoothing scales is due to the fact that the noise level in the reconstructed lensing mass maps is high, and for small smoothing windows, many of the peaks and structures we see in mass maps are the result of noise, which is consistent with the description given in Section 3.3. However, the predicted maps $\kappa_{\rm gal}$, based on real galaxies that have been detected by CFHTLenS, have a lower noise level than the lensing maps. The sources of noise in the predicted maps are likely dominated by intrinsic stochastic biasing between dark matter and baryons and the undetected galaxies in the survey. For this reason, we adopt the predicted maps as the reference from which the peaks will be detected and then compared to the lensing maps. Using the lensing maps as the reference is formally equivalent, but we found that the level of noise in the comparison is reduced when the predicted maps are used instead.
%
%

Next, we want to compare the 2D spatial distribution of peaks between the maps. The peak distribution is a powerful tool that helps visually identify the large scales structures. We will see that this comparison reveals the existence of large underdensities (voids) that cannot be identified with a statistical analysis using moments. Given that for a fixed smoothing scale, the noise in lensing maps is higher than the noise the predicted maps, we decided to detect peaks in the predicted map using the $1.8$ arcmin smoothing scale and compare it to the lensing map using the smoothing scale of $8.9$ arcmin.
A peak location is defined as a pixel where all surrounding pixels have a lower amplitude.
Figure \ref{W1peaks} shows the location of $\kappa_{\rm gal}$ peaks for W1 (shown as white circles) superimposed on the reconstructed lensing map shown as the continuous coloured background. Contours are shown for the lensing reconstruction map at 1,2,3, and 4 sigma levels, which is a common way of indicating the significance of structures in lensing maps.
On average, the distribution of $\kappa_{\rm gal}$ peaks matches the lensing mass overdensities. A quantitative comparison between the predicted convergence and the lensing convergence is shown in the left panel of Figure \ref{baryonspeaks}; the small dots in this figure show, for each peak detected on the $\kappa_{\rm gal}$ map, the corresponding value of the lensing map $\kappa_{\rm obs}$ at the same location. Note that for Figure \ref{baryonspeaks} we have used a smoothing scale of $1.8$ arcmin for the lensing map as well, hence the high noise rms for the lensing peak amplitude. Although only peaks have been used in this plot, it does not mean that the lensing convergence is positive. In fact, there are a substantial number of $\kappa_{\rm gal}$ peaks inside low density regions with negative convergence, despite the fact that, on average, peaks live in positive convergence regions. A comparison of the binned $\kappa_{\rm gal}$ peak values to the averaged lensing convergence in the bin shows that the reconstructed and predicted convergence are strongly correlated, consistent with the fact that baryons trace dark matter to first approximation. The lensing convergence is roughly $2-3$ times lower than the predicted convergence; this is expected, as discussed in Section 4.4, because we assumed that all galaxies are central galaxies of the host dark matter halo. Following the numerical calculations from \cite{2011ApJ...738...45L}, this erroneous assumption would lead to an overestimate of the predicted mass by a factor of $2-3$. Note that for an individual peak, the noise has rms $\sigma_\kappa\simeq 0.015$, consistent with the fact that there are approximately $\sim 100$ lensed galaxies within each smoothing window. Figure \ref{baryonspeaks} also shows the peak analysis performed with the $B$-mode lensing map $\kappa_\perp$, and one can see that the correlation between $\kappa_\perp$ and $\kappa_{\rm gal}$ peaks and troughs vanishes.

\begin{figure*}
 \begin{center}
   \includegraphics[width=\textwidth]{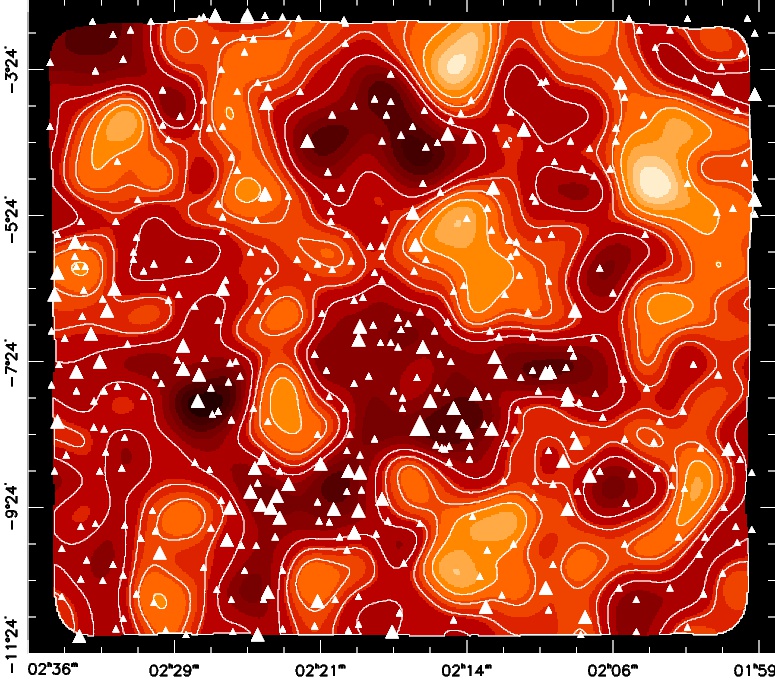}
   \caption{Similar to Figure \ref{W1peaks}, with the background map and contours mapping the projected matter reconstructed from gravitational lensing mass. The open triangles show the position of the troughs in the mass map predicted from the galaxy distribution. The open triangles unambiguously trace the underdense regions in the density mass map. The triangle size is proportional to the height of the trough.}
\label{W1voids}
\end{center}
\end{figure*}

The analysis of the 2D distribution of peaks provides a consistent picture where the galaxy distribution is correlated with the lensing mass, as one would expect.
We can perform a similar analysis on troughs (local minima). A trough location is defined as a pixel where all surrounding pixels have a higher amplitude. A high number density of troughs would be an indication of cosmic voids. Figure \ref{W1voids} shows such an analysis for W1, where white triangles represent troughs in $\kappa_{\rm gal}$. One can see that the triangles preferentially populate underdense regions of the lensing map extending over a few degrees; this is a convincing illustration of the detection of giant projected voids in large scale mass maps based on lensing data. A comparison of Figures \ref{W1peaks} and \ref{W1voids} reveals how the distribution of peaks and troughs provides a mapping of large scale structures and voids, respectively. The right panel on Figure \ref{baryonspeaks} shows how $\kappa_{\rm gal}$ troughs compare to the lensing mass reconstruction $\kappa_{\rm obs}$, with troughs populating large void regions preferentially. This shows that the detection of voids is significant, with residual systematics also being consistent with zero in the underdense regions. Troughs are located primarily in negative convergence regions, as one would expect. The predicted convergence $\kappa_{\rm gal}$ is closer to the reconstructed convergence $\kappa_{\rm obs}$ for the troughs than for the peaks (Figure \ref{baryonspeaks}). A possible explanation is that there are less satellite galaxies in voids than in overdense regions.

\begin{figure*}
 \begin{center}
   \includegraphics[width=\textwidth]{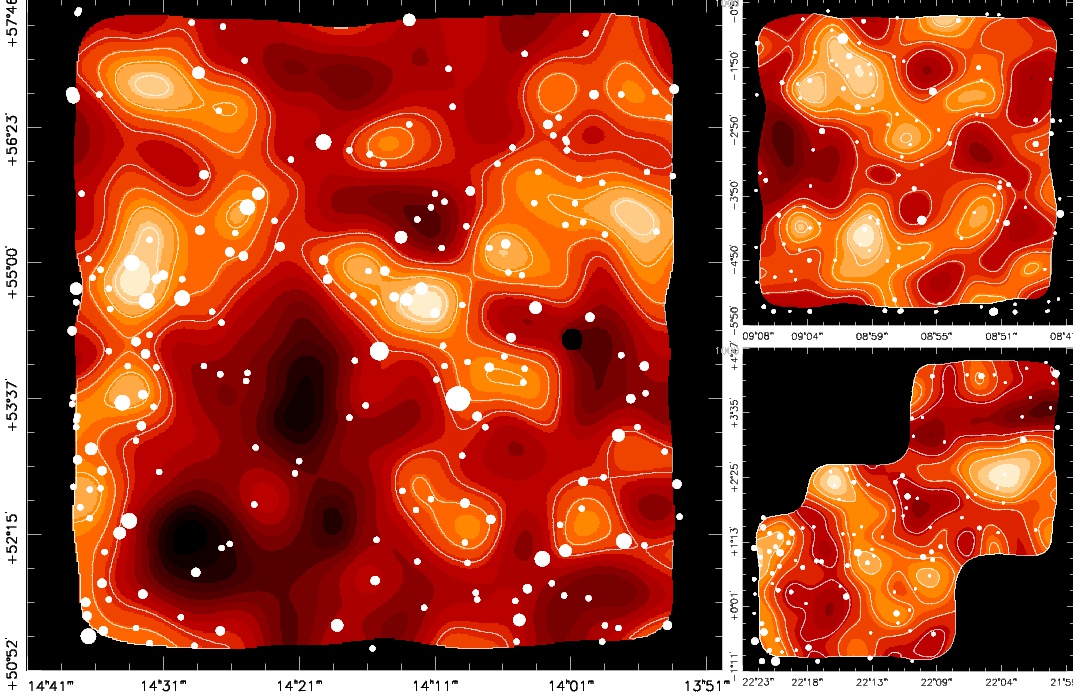}
   \caption{Same as Figure \ref{W1peaks} for the mosaics W3 (left), W2 (top-right), and W3 (bottom-right). The field of view is approximately $7.3\times 7$ deg$^2$, $5\times 5$ deg$^2$, and $6\times 6$ deg$^2$ respectively.}
\label{W234peaks}
\end{center}
\end{figure*}

\begin{figure*}
 \begin{center}
   \includegraphics[width=\textwidth]{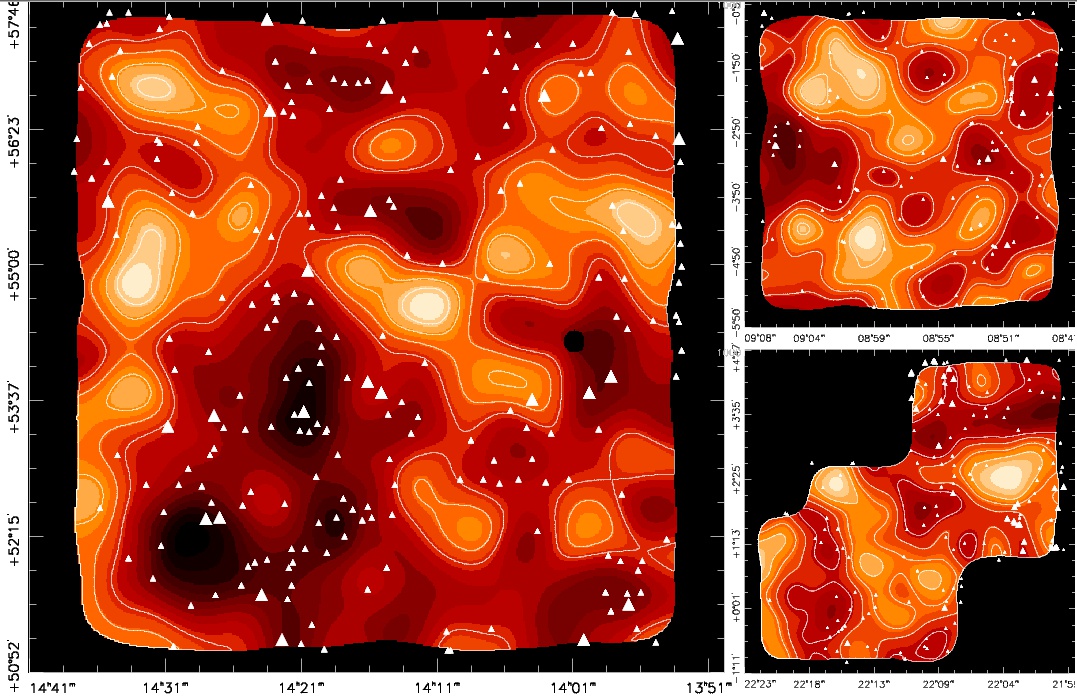}
   \caption{Same as Figure \ref{W1voids} for the mosaics W3 (left), W2 (top-right), and W3 (bottom-right). The field of view is approximately $7.3\times 7$ deg$^2$, $5\times 5$ deg$^2$, and $6\times 6$ deg$^2$ respectively.}
\label{W234voids}
\end{center}
\end{figure*}

Figures \ref{W234peaks} and \ref{W234voids} show the same analysis for the three other CFHTLenS fields W2, W3, and W4. The peak and void statistics shown in Figure \ref{baryonspeaks} include all four fields. The key message from the comparison of the baryon and dark matter maps is that dark matter maps trace large scale structures reliably, but evaluating the significance of most of the individual mass peaks is challenging for two reasons; first, the shot noise is high and very few peaks are above the secure $4\sigma$ threshold. Second, the maps clearly show that the large scale structures are a source of noise comparable to the shot noise \citep{2001A&A...370..743H,2011MNRAS.412.2095H}.

\section{Conclusion}

This paper is the first quantitative cosmological analysis of mass maps reconstructed from lensed galaxy shapes. We validated our approach by using N-body simulations and then applied it to the CFHTLenS data. We find that convergence maps contain reliable cosmological information that has the potential to go beyond the traditional analysis using 2-point statistics. Peak statistics and morphological analysis are the next studies to be performed on mass maps.

Using N-body simulations, we have shown that the reconstruction process is stable and that the input cosmological signal is recovered accurately despite the presence of masks, the relatively high level of shape noise, and the non-Poissonian spatial distribution of the background sources. Mass reconstruction was performed and tested with the traditional KS93 algorithm \citep{1993ApJ...404..441K}. The testing of the full non-linear mass reconstruction at the same level of precision is left for a future study.

The application to the CFHTLenS data shows for the first time that windowed statistics of the convergence can be measured on mass maps. We find excellent agreement between the second and third order moments measured on the reconstructed mass maps and the predictions for a cosmological model determined from a shear correlation function analysis of the same data \citep{2012arXiv1212.3327B,IIGIGG,2013MNRAS.tmp..735K,2013MNRAS.429.2249S}. Our attempt at measuring higher order statistics shows a marginal detection of the fourth order moment. For all the moments of the convergence, the residual systematics are found to be consistent with zero. In Section 4.4, we compared the reconstructed convergence with the predicted convergence using galaxies as tracers of dark matter haloes, where a halo is assigned to each galaxy. We have shown that the predicted and reconstructed mass maps are strongly correlated with each other. The maps reveal the existence of large voids in the projected dark matter distribution, which span regions as large as $3-4$ degrees on the sky.

We consider this paper to be a feasibility study that strongly suggests that future precision cosmology on mass maps is possible. Cosmology with mass maps enables studies that are currently not possible with only the shear (shape) information: global structure morphology, peak statistics with long wavelength modes included, cross correlation with other cosmology maps, and statistics of convergence probability distribution functions. However, before mass maps can become a completely reliable cosmological probe, several technical issues must be addressed:

\begin{enumerate}

\item A correct non-linear reconstruction will have to be implemented to account for the most massive structures, such as clusters of galaxies.

\item Our approach consisted of removing the noise bias from the moments measured on the reconstructed mass. A completely different strategy is to de-noise the map itself before any measurement \citep{2006A&A...451.1139S}. It remains to be tested whether the latter is a robust and better approach for precision measurements.

\item The photometric redshift uncertainty will have to be included in the comparison of dark matter and baryons and in the map making process itself. How to construct a predicted mass map from a distribution of galaxies with photometric redshift errors is an open problem.
    
\item A potential fundamental limitation with maps is that the current mass reconstruction process does not distinguish between intrinsic galaxy alignment and gravitational lensing, which both lead to correlated galaxy shapes. From our particular choice of redshift range, and from the residual systematics studies in CFHTLenS companion papers, we know that our measured signal is mainly caused by gravitational lensing \citep{Heymans2013,Kitching2013}, but future precision cosmology using mass maps will require  a clear identification and separation of the different causes leading to galaxy shapes correlations. It is not clear whether this is possible with mass maps, and further theoretical studies will be needed.
\end{enumerate}

\section*{Acknowledgments}

We would like to thank P.~Schneider and B. Gillis
for insightful comments on the manuscript, and Jennifer Flood for a careful editorial
work on the manuscript. This work is based on
observations obtained with MegaPrime/MegaCam, a joint project of the
Canada-France-Hawaii Telescope (CFHT) and CEA/Irfu, at CFHT, which is
operated by the National Research Council (NRC) of Canada, the
Institut National des Sciences de l'Univers (INSU) at the Centre
National de la Recherche Scientifique (CNRS) of France, and the
University of Hawaii. This research used the facilities of the
Canadian Astronomy Data Centre operated by the NRC of Canada with the
support of the Canadian Space Agency. We thank the CFHT staff, in
particular J.-C.~Cuillandre and E.~Magnier, for the observations, data
processing and continuous improvement of the instrument
calibration. We also thank TERAPIX for quality assessment, and
E.~Bertin for developing some of the software used in this
study. CFHTLenS data processing was made possible thanks to support
from the Natural Sciences and Engineering Research Council of Canada
(NSERC) and HPC specialist O.~Toader. The N-body simulations were
performed on the TCS supercomputer at the SciNet HPC Consortium. The
early stages of the CFHTLenS project were made possible thanks to the
European Commissions Marie Curie Research Training
Network DUEL (MRTN-CT-2006-036133) and its support of CFHTLenS team members LF, HHi, and BR.

\small
The N-body simulations used in this analysis were performed on the TCS supercomputer at the SciNet HPC Consortium. SciNet is funded by: the Canada Foundation for Innovation under the auspices of Compute Canada; the Government of Ontario; Ontario Research Fund - Research Excellence; and the University of Toronto.
\small

LF acknowledges support from NSFC grants 11103012 \& 10878003, Innovation
Program 12ZZ134 and Chen Guang project 10CG46 of SMEC, and STCSM grant
11290706600 \& Pujiang Program 12PJ1406700. CH and FS acknowledge
support from the European Research Council (ERC) through grant
240185. TE is supported by the DFG through project ER 327/3-1 and the
Transregional Collaborative Research Centre TR 33. HHo acknowledges
support from Marie Curie IRG grant 230924, the Netherlands
Organisation for Scientific Research (NWO) through grant 639.042.814
and from the ERC through grant 279396. HHi is supported by the Marie
Curie IOF 252760 and by a CITA National Fellowship. TDK is supported
by a Royal Society University Research Fellowship. YM acknowledges
support from CNRS/INSU and the Programme National Galaxies et
Cosmologie (PNCG). LVW and MJH acknowledge support from NSERC. LVW
also acknowledges support from the Canadian Institute for Advanced
Research (CIfAR, Cosmology and Gravity program). BR acknowledges
support from the ERC through grant 24067, and the Jet Propulsion
Laboratory, California Institute of Technology (NASA). TS acknowledges
support from NSF through grant AST-0444059-001, SAO through grant
GO0-11147A, and NWO.  ES acknowledges support from the NWO grant
639.042.814 and support from ERC under grant 279396. MV acknowledges
support from NWO and from the Beecroft Institute for Particle
Astrophysics and Cosmology.

{\small Author Contributions: All authors contributed to the development and writing of this paper.  The authorship list reflects the lead author of this paper (LVW) followed by two alphabetical groups.  The first alphabetical group includes key contributors to the science analysis and interpretation in this paper, the founding core team and those whose long-term significant effort produced the final CFHTLenS data product.  The second group covers members of the CFHTLenS team who made a significant contribution to either the project, this paper, or both.  The CFHTLenS collaboration was co-led by CH and LVW.}

\normalsize

\section{Appendix}
In this Appendix, we derive the relation between the de-noised moments $\langle\kappa^n_{\rm E}\rangle_{\theta_0}$, $\langle\kappa^n_{\rm B}\rangle_{\theta_0}$ and the observed moments $\langle\kappa^n_{\rm obs}\rangle_{\theta_0}$, $\langle\kappa^n_\perp\rangle_{\theta_0}$, where $\theta_0$ is the Gaussian smoothing scale.
The observed convergence map $\kappa_{\rm obs}$ is the sum of the true signal $\kappa_E$ and an uncorrelated noise component $\kappa_{\rm ran}$. We are repeating Eqs.(\ref{Kobs}) and (\ref{Kperp}):

\begin{equation}
\kappa_{\rm obs}=\kappa_E+\kappa_{\rm ran}.
\end{equation}
The observed $B$-mode convergence map $\kappa_\perp$, which is the reconstructed mass from the $45$ degree rotated galaxies, is similarly related to the true $B$-mode $\kappa_B$ and the noise contribution $\kappa_{\rm ran}$:

\begin{equation}
\kappa_\perp=\kappa_B+\kappa_{\rm ran}.
\end{equation}
Note that the true $B$-mode should be zero if there is no residual $B$-type systematics. Since we are testing this hypothesis precisely, we keep $\kappa_B$ as a quantity to be measured from the data instead of making it equal to zero. The noise contribution to each convergence moment measured in this work are then given by:

\begin{eqnarray}
&&\langle \kappa_{\rm B}^2\rangle_{\theta_0}=\left(\kappa_\perp^2\right)_{\theta_0}-\overline{\left(\kappa_{\rm ran}^2\right)}_{\theta_0}\cr
&&\langle \kappa_{\rm B}^3\rangle_{\theta_0}= \left(\kappa_\perp^3\right)_{\theta_0} \cr
&&\langle \kappa_{\rm E}\kappa_{\rm B}^2\rangle_{\theta_0}= \left(\kappa_{\rm obs}\kappa_\perp^2\right)_{\theta_0} \cr
&&\langle \kappa_{\rm E}^2\kappa_{\rm B}\rangle_{\theta_0}= \left(\kappa^2_{\rm obs}\kappa_\perp\right)_{\theta_0} \nonumber
\end{eqnarray}
\begin{eqnarray}
&&\langle \kappa_{\rm B}^4\rangle_{\theta_0}= \left(\kappa_\perp^4\right)_{\theta_0}
-6\overline{\left(\kappa_\perp^2 \kappa_{\rm ran}^2\right)}_{\theta_0} -\overline{\left(\kappa_{\rm ran}^4\right)}_{\theta_0}\cr
&&\langle \kappa_{\rm E}^3\kappa_{\rm B}\rangle_{\theta_0}= \left(\kappa^3_{\rm obs}\kappa_\perp\right)_{\theta_0} \cr
&&\langle \kappa_{\rm E}^2\kappa_{\rm B}^2\rangle_{\theta_0}= \left(\kappa^2_{\rm obs}\kappa_\perp^2\right)_{\theta_0}
-\overline{\left(\kappa^2_{\rm obs}\kappa_{\rm ran}^2\right)}_{\theta_0} -\overline{\left(\kappa_\perp^2\kappa_{\rm ran}^2\right)}_{\theta_0}\cr
&&-\overline{\left(\kappa_{\rm ran}^4\right)}_{\theta_0}\cr
&&\langle \kappa_{\rm E}\kappa_{\rm B}^3\rangle_{\theta_0}= \left(\kappa_{\rm obs}\kappa_\perp^3\right)_{\theta_0} \nonumber
\end{eqnarray}
\begin{eqnarray}
&&\langle \kappa_{\rm B}^5\rangle_{\theta_0}= \left(\kappa^5_{\perp}\right)_{\theta_0}-10\overline{\left(\kappa_\perp^3\kappa_{\rm ran}^2\right)}_{\theta_0}\cr
&&\langle \kappa_{\rm E}\kappa_{\rm B}^4\rangle_{\theta_0}= \left(\kappa_{\rm obs}\kappa_\perp^4\right)_{\theta_0}\cr
&&\langle \kappa_{\rm E}^2\kappa_{\rm B}^3\rangle_{\theta_0}= \left(\kappa^2_{\rm obs}\kappa_\perp^3\right)_{\theta_0}\cr
&&\langle \kappa_{\rm E}^3\kappa_{\rm B}^2\rangle_{\theta_0}= \left(\kappa^3_{\rm obs}\kappa_\perp^2\right)_{\theta_0}-\overline{\left(\kappa^3_{\rm obs}\kappa_{\rm ran}^2\right)}_{\theta_0}\cr
&&\langle \kappa_{\rm E}^4\kappa_{\rm B}\rangle_{\theta_0}= \left(\kappa^4_{\rm obs}\kappa_\perp\right)_{\theta_0}
\label{noisemoments}
\end{eqnarray}
The left-hand side is the de-noised (unbiased) moment and the right-hand side shows the combination of moments leading to this unbiased measurement. The measured signal, $B$-mode, and random noise maps are given by $\kappa_{\rm obs}$, $\kappa_\perp$, and $\kappa_{\rm ran}$ respectively. As a reminder from the definition given in Section 4.2, a few ensemble averages have been introduced: $\left(...\right)_{\theta_0}$ denotes the moment average over {\it one} map, $\overline{\left(...\right)}_{\theta_0}$ is averaged over several noise maps $\kappa_{\rm ran}$, and $\langle...\rangle_{\theta_0}$ is the de-noised moment.
One can see from these equations that there are as many systematics moments as there are permutations of $\kappa_E$ and $\kappa_B$ for a given moment order. A total of one hundred random maps have been generated for each smoothing scale. In order to derive Eqs.(\ref{noisemoments}), we assumed that all the terms with odd order moments of the noise maps $\kappa_{\rm ran}$ can be neglected, as they only introduce noise in the equation above.

\bibliographystyle{mn2e_mod}

\bibliography{lensing}

\label{lastpage}
\end{document}